\documentclass[12pt]{article}
\usepackage{graphicx}

\newcommand\pubnumber{NuPhys2017-Masip}
\newcommand\pubdate{\today}

\def\napoli{Departamento de F\'\i sica Te\'orica y del Cosmos\\
Universidad de Granada, E-18071 Granada, SPAIN}
\def\support{\footnote{Work supported by the MINEICO of Spain under grant
FPA2016-78220-C3-2-P}}

\def\Title#1{\begin{center} {\Large #1 } \end{center}}
\def\Author#1{\begin{center}{ \sc #1} \end{center}}
\def\Address#1{\begin{center}{ \it #1} \end{center}}

\newcommand\pubblock{\rightline{\begin{tabular}{l} \pubnumber\\
         \pubdate  \end{tabular}}}
\newenvironment{Abstract}{\begin{quotation}  }{\end{quotation}}
\newenvironment{Presented}{\begin{quotation} \begin{center} 
             PRESENTED AT\end{center}\bigskip 
      \begin{center}\begin{large}}{\end{large}\end{center} \end{quotation}}
\def\Acknowledgements{\bigskip  \bigskip \begin{center} \begin{large}
             \bf ACKNOWLEDGEMENTS \end{large}\end{center}}
             \newcommand{\beq}{\begin{equation}}
\def\beq{\begin{equation}}
\def\eeq#1{\label{#1}\end{equation}}
\def\eeqn{\end{equation}}

\def\beqa{\begin{eqnarray}}
\def\eeqa#1{\label{#1}\end{eqnarray}}
\def\eeqan{\end{eqnarray}}

\let\bar=\overbar

\def\Dslash{\not{\hbox{\kern-4pt $D$}}}
\def\dslash{\not{\hbox{\kern-2pt $\del$}}}

\def\msb{{\bar{\ssstyle M \kern -1pt S}}}

\begin{document}
\begin{titlepage}
\pubblock

\vfill
\Title{High energy neutrinos}
\vfill
\Author{Manuel Masip\support}
\Address{\napoli}
\vfill
\begin{Abstract}
We describe several components in the diffuse flux of high energy neutrinos  
reaching the Earth
and discuss whether they could explain IceCube's observations. Then we focus on
TeV neutrinos from the Sun. We show that this solar $\nu$ flux is 
correlated with the cosmic-ray shadow of the Sun measured by HAWC, and we find
that it is much larger than the flux of atmospheric neutrinos. Stars like our Sun provide
neutrinos with a very steep spectrum and no associated gammas. We argue 
that this is the type of contribution that could solve the main puzzle presented by the 
high energy IceCube data.
\end{Abstract}
\vfill
\begin{Presented}
NuPhys2017, Prospects in Neutrino Physics\\
Barbican Centre, London, UK, December 20--22,  2017
\end{Presented}
\vfill
\end{titlepage}
\def\thefootnote{\fnsymbol{footnote}}
\setcounter{footnote}{0}

\section{Introduction}

IceCube observations \cite{Aartsen:2013jdh} are the 
main reason why high energy neutrinos are right now specially interesting.
IceCube has shown that neutrinos with energy up to several thousand TeV
are there and that they can be detected and studied. Eventually, we will learn 
about their interations at these huge energies. We may use them, for 
example, to put bounds
on the mass of leptoquarks \cite{Anchordoqui:2006wc} or on 
TeV gravity models \cite{Illana:2014bda}. And with them
we will also learn Astrophysics, as we
will answer the basic question  {\it Where do these neutrinos come from?} 
that is actually the main topic of this talk.

First I will briefly review the possible contribution to the IceCube signal
of several components present in the
diffuse flux of neutrinos reaching the Earth. Then I will discuss a source of TeV neutrinos
that recently has attracted renewed attention, the Sun. Finally I will argue that 
the data may be suggesting an unexpected scheme for the origin of the IceCube 
neutrinos.

\section{Components in the diffuse flux of neutrinos}
Neutrinos of energy above 1 GeV have always a hadronic origin, 
they are secondary particles created in the collisions of high energy cosmic rays
(CRs) with matter or light. These collisions occur as CRs enter the atmosphere
and generate atmospheric neutrinos or wherever we find CRs and matter. In
particular, in the interstellar (IS) and intergalactic (intracluster) space, where certainly there
are CRs and also plenty of gas.

It is then clear that the key to understand any neutrino flux is the parent CR flux.
At energies below $E_{\rm knee}\approx 10^{6.5}$ GeV
we find that the CR flux reaching the Earth is dominated by hydrogen and He, 
with fluxes
[in particles/(GeV sr s cm$^2$)]:
\begin{equation}
\Phi_p = 1.3 \left( {E\over {\rm GeV}} \right)^{-2.7},
\hspace{1cm}
\Phi_{\rm He} = 0.54 \left( {E\over {\rm GeV}} \right)^{-2.6}.
\label{fluxL}
\end{equation}
At higher energies up to $E_{\rm ankle}\approx 10^{9.5}$ GeV 
the spectral index changes to $\alpha\approx 3$,
\begin{equation}
\Phi = 330 \left( {E\over {\rm GeV}} \right)^{-3.0}\,,
\label{fluxH}
\end{equation}
and the composition is uncertain.
This flux can be understood within the following basic scheme. 
After they are accelerated according to a power law
$E^{-\alpha_0}$, galactic CRs {\it diffuse} from the sources and 
stay trapped by magnetic fields during a time proportional 
to $E^{-\delta}$. As a consequence, the spectral index that we
see is $\alpha=\alpha_0+\delta$, 
reflecting that higher energies
are less frequent both because CRs are produced at a lower 
rate and because they propagate 
with a larger diffusion coefficient and 
leave our galaxy faster. The transport parameter
$\delta$ is universal (identical for CRs with the same rigidity
$R=E/Ze$) and its value, determined by
the spectrum of magnetic turbulences in the IS medium, 
may be constant up to $E_{\rm ankle}$. This scheme suggests 
{\it (i)} that the CR density 
will be a factor of $(B/B_0)^\delta$ larger 
in galactic regions where the mean magnetic field strength is larger
and {\it (ii)} that (assuming a steady state) our galaxy emits into the intracluster 
space CRs with a $E^{-(\alpha-\delta)}$ spectrum harder than
the one we see at the Earth.
We can then discuss the neutrinos produced by these CRs in the different 
environments. 

\begin{itemize}

\item The atmospheric neutrino flux includes two components: the so called conventional
neutrinos from $\pi$ and $K$ decays \cite{Lipari:1993hd}, and the $\nu$ flux from 
charmed hadron decays \cite{Halzen:2016thi}. At TeV
energies light mesons tend to collide before they decay, which increases the spectral
index of conventional neutrinos in a unit (see Fig.~\ref{fig1}). 
Neutrinos from charm inherit the index
from the parent CRs and dominate the atmospheric flux at $E>200$--$300$ TeV.

%%%%%%%%%%%%%%%%%%%%%%%%%%%%%%%%%%%%%%%%%%%%%%%%%%%%%%%%%%%%%%%%%%%%%%%%%
%%
%%   use this format to include an .eps figure into your paper
%%
\begin{figure}[t]
\centering
\includegraphics[height=2.0in]{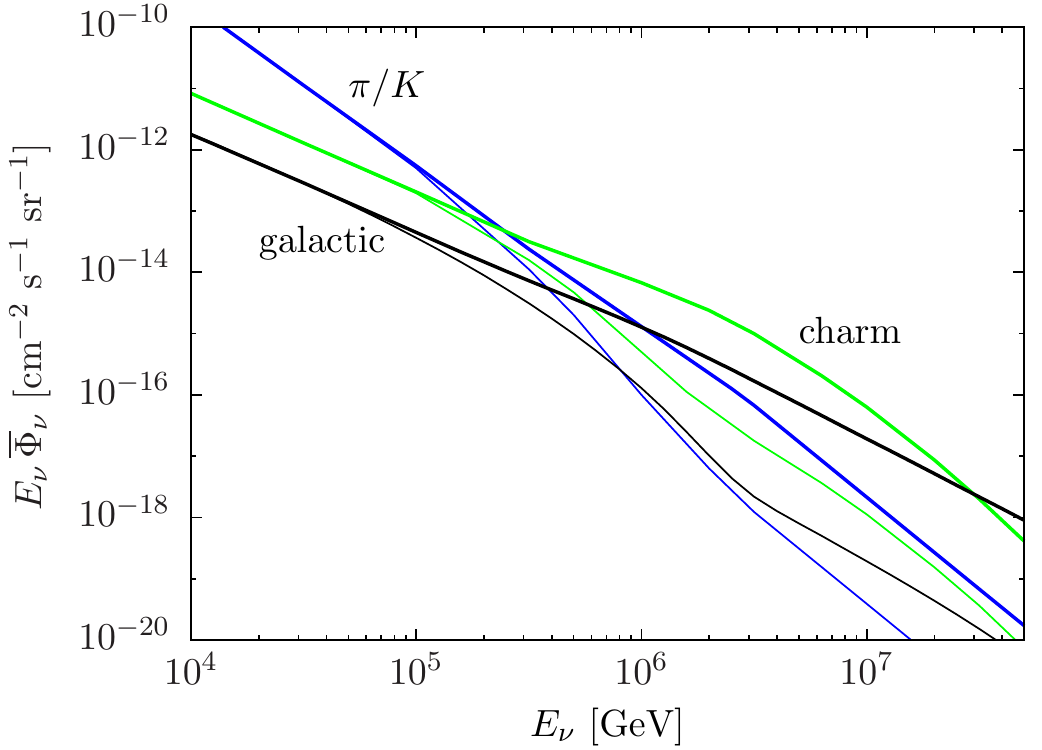}\hspace{0.7cm}
\includegraphics[height=1.96in]{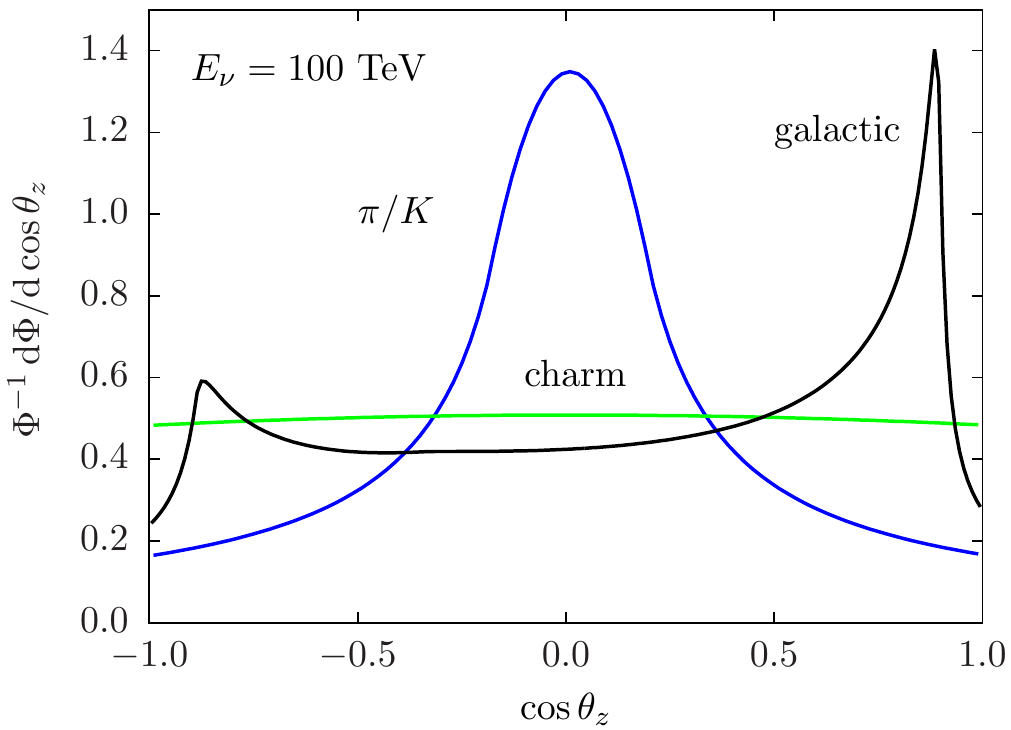}
\caption{Atmospheric and galactic $\nu$ fluxes and their zenith distribution at IceCube.}
\label{fig1}
\end{figure}
%%%%%%%%%%%%%%%%%%%%%%%%%%%%%%%%%%%%%%%%%%%%%%%%%%%%%%%%%%%%%%%%%%%%%%%%%%%

\item The diffuse flux of galactic neutrinos from CR collisions in the IS space 
($\Phi_{\rm gal}$) is 
mostly distributed near the galactic plane \cite{Stecker:1978ah}. 
We see in Fig.~\ref{fig1} that $\Phi_{\rm gal}$ is
just a 25\% correction to the atmospheric $\nu$ flux from charm, having both fluxes 
a similar spectrum \cite{Carceller:2016upo}. The lines at high energies reflect the uncertainty in the CR
composition at $E\ge E_{\rm knee}$.
We also plot the zenith distribution of these fluxes at IceCube. Notice
that the conventional $\nu$ flux, which dominates below 100 PeV, is 7 times larger from
near horizontal than from vertical directions.

\item CR collisions with gas 
also happen in the IS medium of other galaxies similar to
ours. We then expect another component in the high-energy 
$\nu$ flux reaching the Earth defined by the ensemble of all other galaxies
($\Phi_{\rm AG}$). Its spectrum will be similar to the one in $\Phi_{\rm gal}$,
but it will be isotropically distributed.

\item The final component ($\Phi_{\rm IG}$) is generated by CR collisions in intracluster 
space \cite{Berezinsky:1996wx}. 
These CRs have a harder spectrum; if we take $\delta\approx 0.5$
their spectral index is around $2.1$ at $E<E_{\rm knee}$ and $2.4$ at higher energies. 
The intergalactic medium is thinner 
than the galactic one, but CRs may spend there a time of the order of the age of the 
universe and the probability of interaction may be not negligible. 
The diffuse $\nu$ flux $\Phi_{\rm IG}$ inherits the hard 
spectrum of the parent CRs.

\end{itemize}

%%%%%%%%%%%%%%%%%%%%%%%%%%%%%%%%%%%%%%%%%%%%%%%%%%%%%%%%%%%%%%%%%%%%%%%%%
%%
%%   use this format to include an .eps figure into your paper
%%
\begin{figure}[tb]
\centering
\includegraphics[height=2.4in]{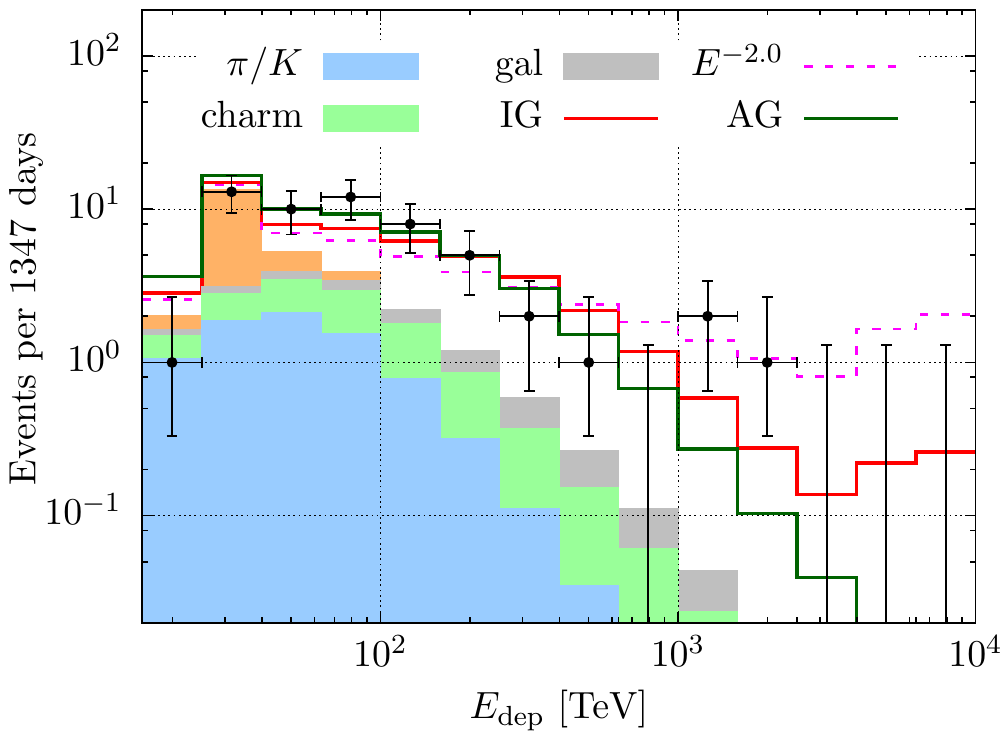}
\caption{Events at IceCube implied by the different neutrino fluxes (see text).}
\label{fig2}
\end{figure}
%%%%%%%%%%%%%%%%%%%%%%%%%%%%%%%%%%%%%%%%%%%%%%%%%%%%%%%%%%%%%%%%%%%%%%%%%%%

Let us take these 5 neutrino fluxes and estimate how many events they imply at
IceCube \cite{Carceller:2017tvc}. 
We will focus on high energy starting events (HESE) observed in
4 years of data \cite{Aartsen:2015zva}.
In Fig.~\ref{fig2} we plot the atmospheric (blue and green) and the
galactic (grey) contributions together with the expected contribution from atmospheric
muons entering the detector from outside (orange), that introduce a large uncertainty 
at energies below 60 TeV. To these events we add an extragalactic 
contribution of type $\Phi_{\rm AG}$, $\Phi_{\rm IG}$ or $\Phi\approx E^{-2.0}$ with a
normalization that optimizes the fit.

We see, first of all, that in order to reproduce the $70$--$700$ TeV  data 
the extragalactic $\nu$ flux that we need to add 
is very steep, even steeper than $\Phi_{\rm AG}\approx E^{-2.6}$. 
The problem with these fluxes
is that they imply too many gamma rays at low energies. 
Neutrinos appear correlated with gammas, 
and the extrapolation of the gamma flux associated to this $\Phi_{\rm AG}$ down
to $10$--$100$ GeV would be inconsistent with the data from Fermi-LAT 
and 
other older observatories \cite{Stecker:1978ah}. 
The flux $\Phi_{\rm AG}$ must then be much smaller and basically negligible
at all IceCube energies: although the data clearly prefers very
steep $\nu$ fluxes, the spectral index should not
be larger than $2.1$ \cite{Murase:2013rfa} to avoid an excess of diffuse gammas at Fermi-LAT energies.

This problem persists after the fifth year of IceCube data (not in the plot), which 
does not include any new events in the higher energy bins. There we find 2, 1, 0, 2 and 1 events, 
not a rich statistics but enough to suggest a much flatter spectrum than at lower
energies. It is apparent that a single power law can not fit IceCube's 
HESE events \cite{Anchordoqui:2016ewn}. 
A harder flux at high energies, however, may have problems
as well. In particular, one should
explain why we see events at $1$--$2$ PeV but not at $6.3$ PeV, where
the Glashow resonance (a $W^-$ boson in the $s$ channel of $\bar \nu_e e$
collisions) gives a large contribution. Whatever the origin of the PeV 
IceCube neutrinos, the absence of the Glashow resonance 
could indicate a change in the composition of the parent CRs towards a higher 
mass number at higher energies.
It is indeed puzzling, and in this context we would like to discuss 
a different but possibly related topic.

\section{TeV neutrinos from the Sun}

High energy CRs may reach the surface of the Sun, shower there and give neutrinos
that may be detected at the Earth \cite{Seckel:1991ffa}. 
Let us discuss briefly the main issues involved in 
the calculation of this solar neutrino flux \cite{Masip:2017gvw}.

%%%%%%%%%%%%%%%%%%%%%%%%%%%%%%%%%%%%%%%%%%%%%%%%%%%%%%%%%%%%%%%%%%%%%%%%%
%%
%%   use this format to include an .eps figure into your paper
%%

\begin{figure}[tb]
\vspace{-1cm}
\begin{minipage}{0.45\linewidth}
\includegraphics[width=2.7in]{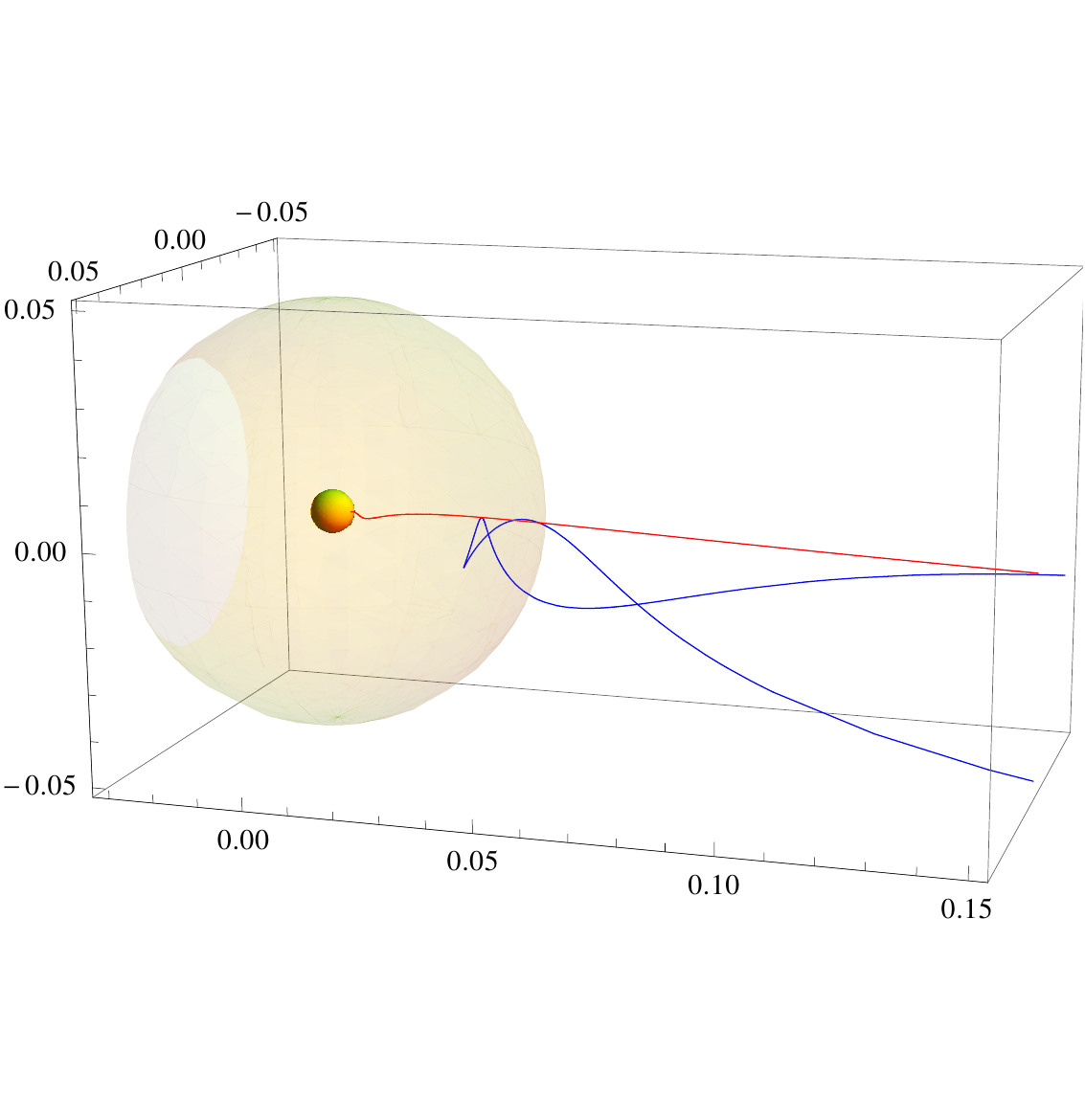}
\end{minipage}
\hspace{0.3cm}
\begin{minipage}{0.45\linewidth}
\includegraphics[width=2.9in]{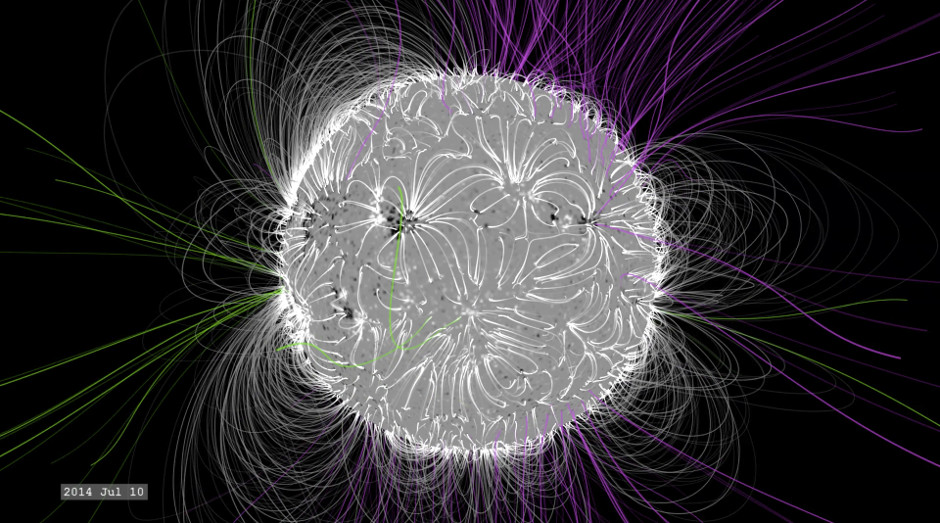}
\end{minipage}
\vspace{-1cm}
\caption{CR trajectories through the Parker field and 
field lines near the Sun surface \cite{solar}.}
\label{fig3}
\end{figure}
%%%%%%%%%%%%%%%%%%%%%%%%%%%%%%%%%%%%%%%%%%%%%%%%%%%%%%%%%%%%%%%%%%%%%%%%%%%

First and probably the most difficult one, can 
CRs really reach the Sun? The question makes sense because the solar magnetic field 
has a radial component that grows like $1/r^2$ when CRs approach the Sun, so they 
may experience a magnetic mirror effect before they reach the surface (see Fig.~\ref{fig3}).
At distances beyond $10R_\odot$ we find the 
Parker (interplanetary) field, but at shorter distances the magnetic structure is 
much more complex.
Field lines tend to corrotate with the Sun, and near the surface there are closed lines that start
and finish in the surface. In addition, this magnetism is not stable, it has a 11 year cycle
correlated with the  solar activity.

\begin{figure}[t]
\centering
\includegraphics[height=2.4in]{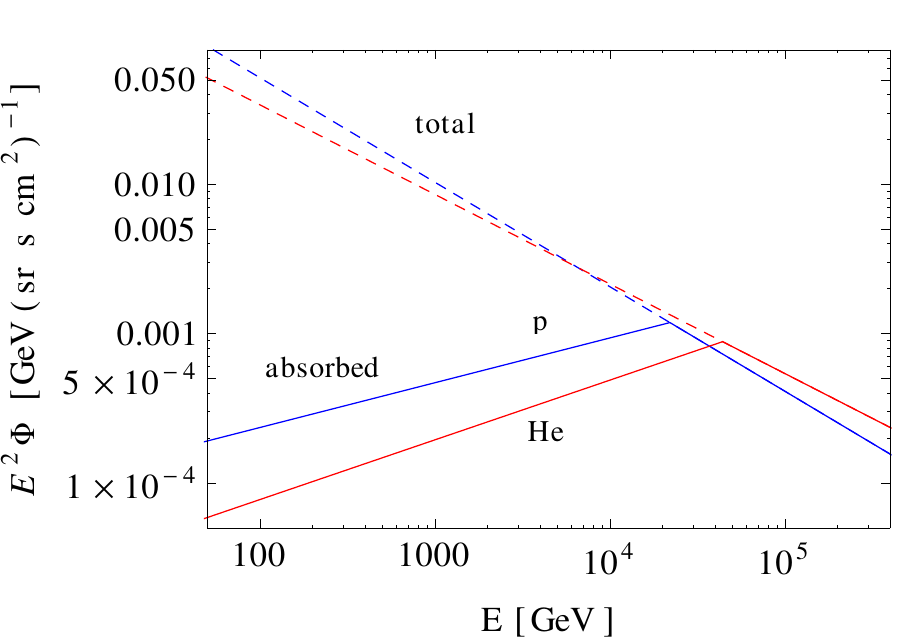}
\caption{Flux of CRs absorbed by the Sun.}
\label{fig4}
\end{figure}

Fortunately, the magnetic effects on CRs are simplified by the fact that the flux
is basically isotropic. The solar field acts like a magnetic lens, and we know from Liouville's 
theorem that a lens (including a mirror) will not make anisotropic an isotropic flux: the
only possible effect of the Sun on the CR flux is then to interrupt trajectories 
that were aiming to the Earth,
{\it i.e.}, to create a shadow. This shadow, first measured by TIBET and more
recently by other observatories, reveals the
absorption rate of CRs by the Sun. HAWC, in particular, 
has studied its energy dependence \cite{Enriquez:2015nva}. The 
CR shadow appears at 2 TeV, and
it is not a black disk but a deficit that decreases radially along an angular distance 
10 times larger than the Sun. If we integrate the deficit we find that
it represents a 6\% of the shadow at 2 TeV, a 27\% at 8 TeV and the complete shadow
at 50 TeV.
This means that at energies below the TeV most CRs are mirrored and
do not reach the  surface, whereas at 50 TeV there is a full set of CR
trajectories that where 
aiming to the Earth but were absorbed by the Sun. In Fig.~\ref{fig4} we plot
a flux of absorbed CRs that coincides with the one we see at high energies but
changes at a given rigidity and reproduces HAWC's observations.

%%%%%%%%%%%%%%%%%%%%%%%%%%%%%%%%%%%%%%%%%%%%%%%%%%%%%%%%%%%%%%%%%%%%%%%%%
%%
%%   use this format to include an .eps figure into your paper
%%
%%%%%%%%%%%%%%%%%%%%%%%%%%%%%%%%%%%%%%%%%%%%%%%%%%%%%%%%%%%%%%%%%%%%%%%%%%%

Once they reach the Sun, CRs face a very thin environment.
The photosphere has a total depth of just 2.7 g/cm$^2$ along 500 km, 
and then the solar suface is not
dense  like in the Earth, it takes 1500 km to cross just 
100 g/cm$^2$ of matter. As a
consequence, high energy pions and kaons produced there 
have plenty of time to decay giving 
leptons before they collide. 
This is in contrast with what happens in the Earth's atmosphere
(of higher density), 
and it is 
the main reason why the high-energy neutrino flux from the Sun is much larger
than the atmospheric one.

\begin{figure}[t]
\centering
\includegraphics[height=2.03in]{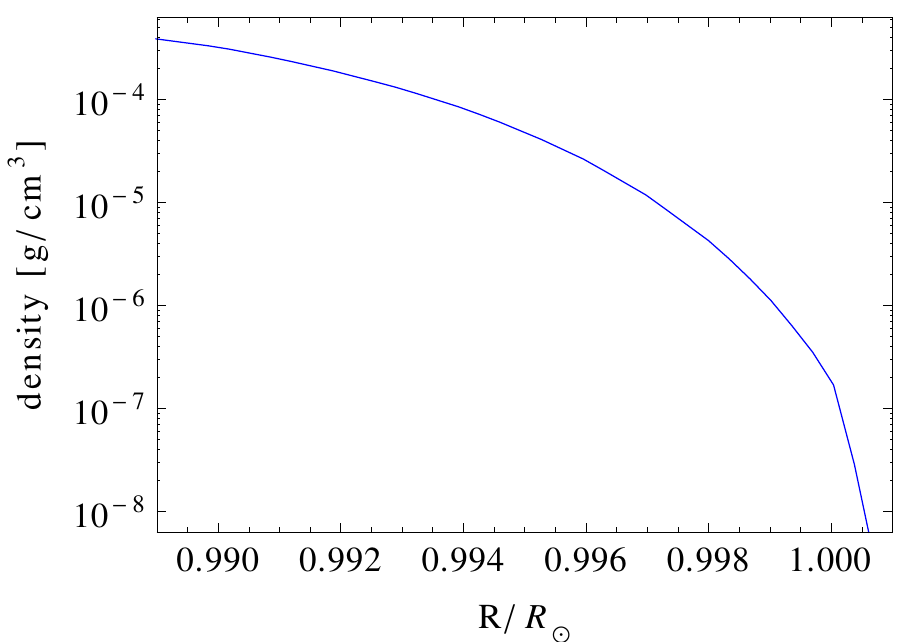}\hspace{0.7cm}
\includegraphics[height=2.0in]{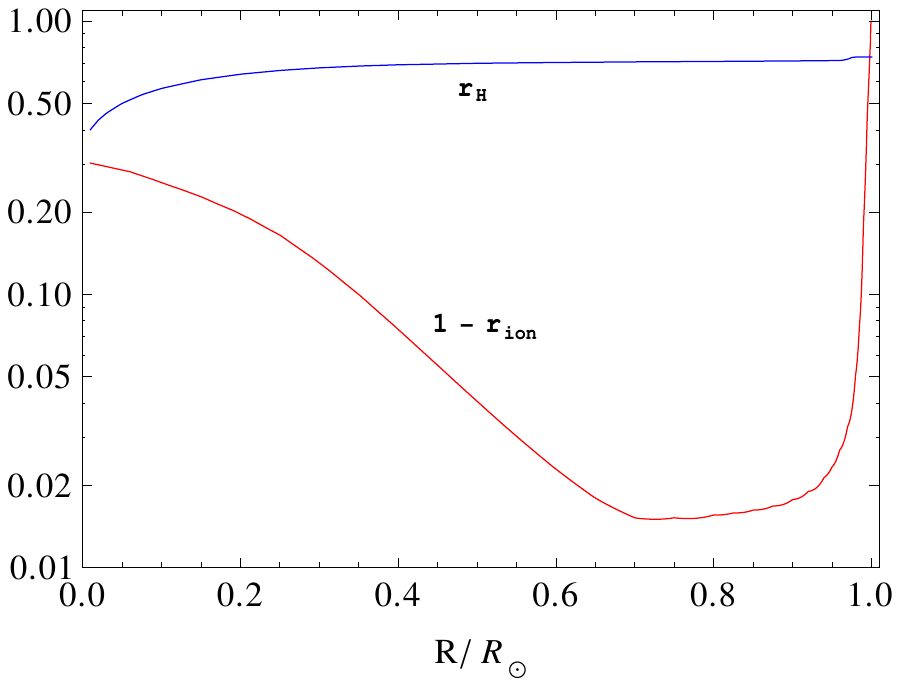}
\caption{Density and fractions of hydrogen
and non-ionized matter in the Sun.}
\label{fig5}
\end{figure}

Another important factor that separates Sun from Earth showers
is the different propagation of
 muons. At high muon energies radiative processes have a smaller cross 
section than in the Earth, as the Sun is composed of elements with lower atomic number 
(H and He). And at low 
energies the loss by ionization is also going to be much smaller there, since
most of the matter is in the Sun is ionized. In Fig.~\ref{fig5} we plot the fraction of hydrogen
(the rest is mostly $^4$He) and the fraction of matter that is not ionized at different
solar radii.

%%%%%%%%%%%%%%%%%%%%%%%%%%%%%%%%%%%%%%%%%%%%%%%%%%%%%%%%%%%%%%%%%%%%%%%%%
%%
%%   use this format to include an .eps figure into your paper
%%
%%%%%%%%%%%%%%%%%%%%%%%%%%%%%%%%%%%%%%%%%%%%%%%%%%%%%%%%%%%%%%%%%%%%%%%%%%%

%%%%%%%%%%%%%%%%%%%%%%%%%%%%%%%%%%%%%%%%%%%%%%%%%%%%%%%%%%%%%%%%%%%%%%%%%
%%
%%   use this format to include an .eps figure into your paper
%%
\begin{figure}[b]
\centering
\vspace{-3.5cm}
\includegraphics[height=4in]{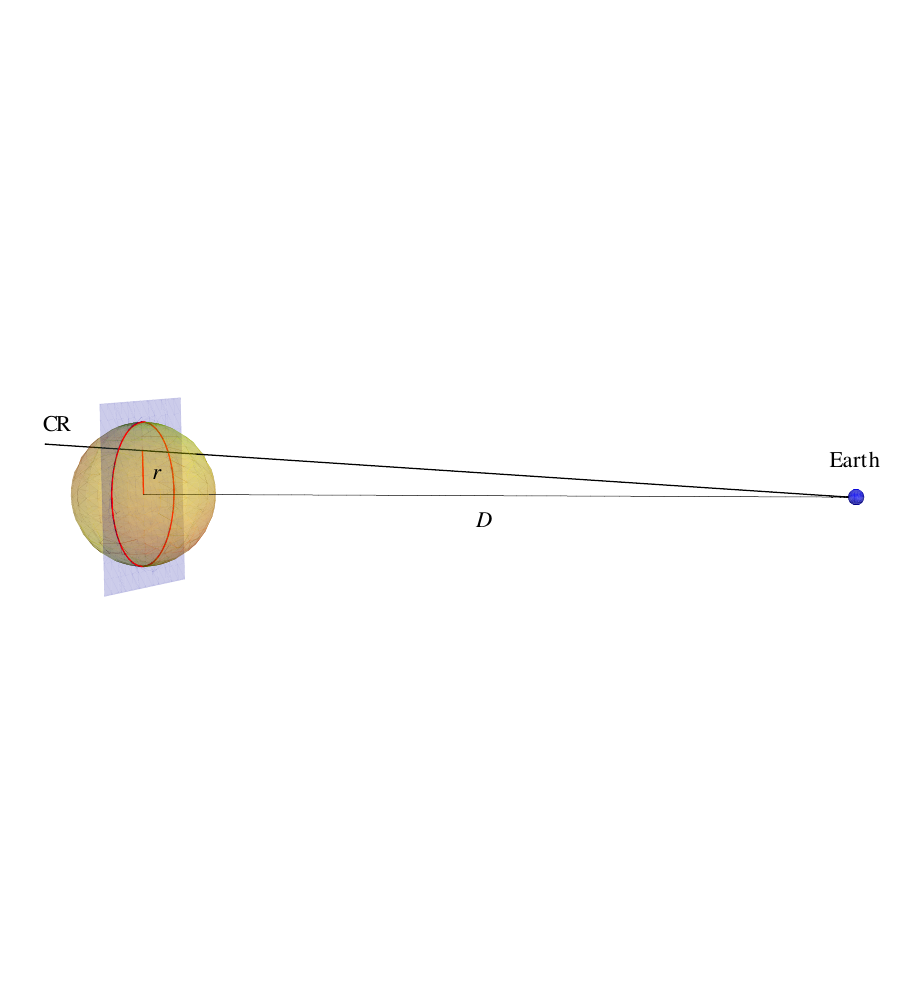}
\vspace{-4cm}
\caption{CR shower at a transverse 
distance $r\le R_\odot$ from the Sun's center.}
\label{fig6}
\end{figure}
%%%%%%%%%%%%%%%%%%%%%%%%%%%%%%%%%%%%%%%%%%%%%%%%%%%%%%%%%%%%%%%%%%%%%%%%%%%

The calculation of the neutrino flux reaching the Earth is then simple. Let us first assume that
the solar shower follows a straight line of transverse parameter $r$, as shown in Fig.~\ref{fig6}.
We can write the transport equations (see details in \cite{Masip:2017gvw}) for the 17 relevant
species along that line: hadrons
($p,n,\bar p,\bar n, \pi^\pm, K^\pm,K_L$), muons  ($\mu_L^\pm, \mu_R^\pm$)
and neutrinos ($\nu_{e,\mu}, \bar \nu_{e,\mu}$). The yields from hadron collisions 
have been obtained with
EPOS-LHC \cite{Pierog:2013ria}, and in the decay yields 
it is important to distinguish between muons of both helicities \cite{Lipari:1993hd}.
It is also straightforward to include neutrino oscillations; the main effect takes place in vacuum,
between the Sun and the Earth. We find that at energies below 5 TeV the averaged oscillations
imply basically the same frequency for the three neutrino flavors at the Earth. A final observation
concerns our assumption of a straight shower unaffected by the solar magnetic field, which
should be good only at 
large energies. At lower energies ($E<1$ TeV), however, all
particles producing neutrinos (both light mesons and muons) decay before losing energy
and the $\nu$ absorption by 
the Sun is negligible. As a consequence, the neutrino yield does not 
depend on the trajectory of the parent particles (the Sun's emission is isotropic) and 
the assumed straight shower  gives also an acceptable approximation.

%%%%%%%%%%%%%%%%%%%%%%%%%%%%%%%%%%%%%%%%%%%%%%%%%%%%%%%%%%%%%%%%%%%%%%%%%
%%
%%   use this format to include an .eps figure into your paper
%%
\begin{figure}[t]
\centering
\includegraphics[height=2.4in]{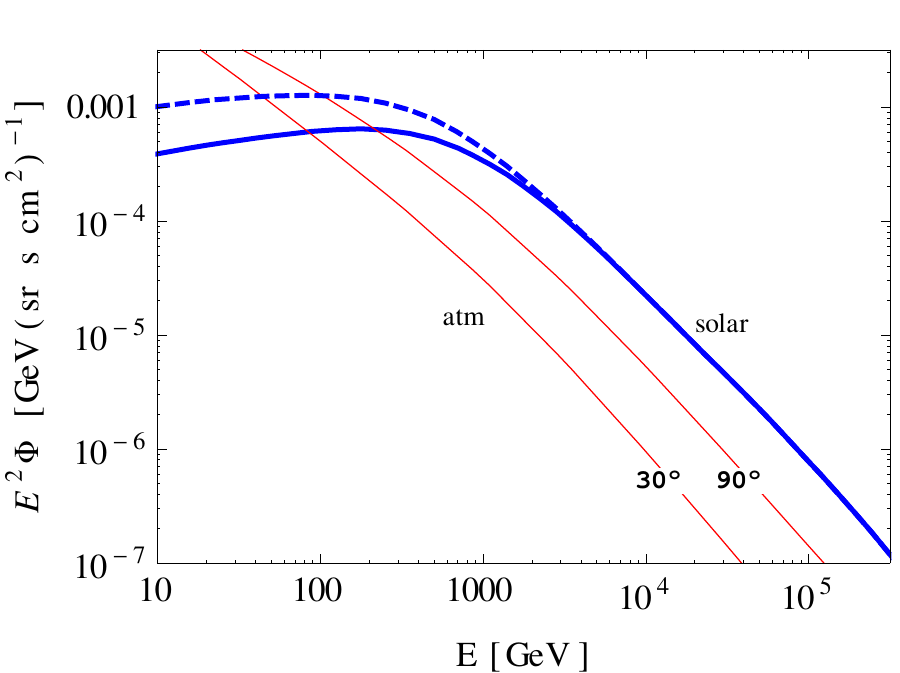}
\caption{Total neutrino fluxes at the Earth.}
\label{fig7}
\end{figure}
%%%%%%%%%%%%%%%%%%%%%%%%%%%%%%%%%%%%%%%%%%%%%%%%%%%%%%%%%%%%%%%%%%%%%%%%%%%

Fig.~\ref{fig7} summarizes our results for 
periods of high (solid) and low (dashes) solar activity. We plot the average $\nu$ flux
over the whole angular region defined by the Sun ($0.27^\circ$ of radius).
The flux is well above the atmospheric $\nu$ flux, specially from vertical directions.
For example, at  $\theta_z=30^\circ,150^\circ$ the $\nu_\mu$ flux from the Sun is 7 times 
larger than the atmospheric one,
and the total flux (adding the three flavors) is 20 times larger. 
At IceCube, where the Sun is always low in the horizon, 
the solar signal would be very difficult to see: an order 1 correction to $\Phi_{\rm atm}$
in just a very small area. KM3NeT, on the other hand, may follow the Sun under 
smaller zenith angles and has a better angular resolution, so it should
be able to detect this solar flux.

\section{Outlook}
The Sun is probably the brightest object in the sky also for high energy neutrinos. The
flux that it emits has a very steep spectrum,
as higher energy neutrinos are partially absorbed by the Sun.
At energies below 1 TeV 
the flux flattens just because the parent CRs find it difficult to reach the solar surface.
At any rate, this flux is a strong background in indirect dark matter searches at
neutrino telescopes.

One may wonder if there is possibly a relation between these solar neutrinos and 
the high-energy IceCube data. Obviously, not a direct relation, but 
we have identified a new source of high energy neutrinos
--stars like our Sun-- that gives  a very steep neutrino flux and, most important, 
that does not produce gamma rays (most of them are unable to 
scape the star once produced). Indeed, IceCube data suggests such a steep 
flux at $E\approx 100$ TeV, which can be reconciled with Fermi-LAT only if 
it does not come together with gammas. A second and much harder 
component, possibly from intracluster CR interactions, should dominate
the neutrino flux at PeV energies.

High energy astroparticles define a puzzle where all the pieces must fit
together: a diffuse flux of neutrinos with no gammas,
or IceCube events at 2 PeV but not at the 6 PeV Glashow resonance, are observations
that must mean something. The puzzle is far from complete, but the prospects are
exciting.

\Acknowledgements
I would like to thank Francesca Di Lodovico, Silvia Pascoli and Peter Ballet for 
their kind invitation to participate in NuPhys2017.

\end{document}